\documentclass{aastex7}
\usepackage{amsmath}

\begin{document}
\title{A Study Revealing Physical Attributes of Supernova Remnant in G321.3-3.9}

\author[orcid=0000-0002-0829-7553,gname=Shaobo, sname=Zhang]{Shaobo Zhang}
\affiliation{National Astronomical Observatories, Chinese Academy of Sciences, Beijing 100012, China}
\affiliation{School of Astronomy and Space Science, University of Chinese Academy of Sciences, Beijing 100049, China}
\email{shbzhang@nao.cas.cn}

\author[gname=Xianhuan, sname=Lei]{Xianhuan Lei}
\affiliation{National Astronomical Observatories, Chinese Academy of Sciences, Beijing 100012, China}
\affiliation{School of Astronomy and Space Science, University of Chinese Academy of Sciences, Beijing 100049, China}
\affiliation{School of Physics and Electronic Science, Guizhou Normal University, Guiyang 550025, China}
\email{202408030@gznu.edu.cn}

\author[orcid=0000-0003-3775-3770,gname=Hui, sname=Zhu]{Hui Zhu}
\affiliation{National Astronomical Observatories, Chinese Academy of Sciences, Beijing 100012, China}
\email[show]{zhuhui@bao.ac.cn}

\author[orcid=0009-0001-9789-3858,gname=Xueying, sname=Hu]{Xueying Hu}
\affiliation{School of Physics and Astronomy, Beijing Normal University, Beijing, 100875, China}
\email{xy_hu@mail.bnu.edu.cn}

\author[orcid=0000-0002-6322-7582,gname=Xiaohong, sname=Cui]{Xiaohong Cui}
\affiliation{National Astronomical Observatories, Chinese Academy of Sciences, Beijing 100012, China}
\email[show]{xhcui@bao.ac.cn}

\author[orcid=0000-0002-9079-7556,gname=Wenwu, sname=Tian]{Wenwu Tian}
\affiliation{National Astronomical Observatories, Chinese Academy of Sciences, Beijing 100012, China}
\affiliation{School of Astronomy and Space Science, University of Chinese Academy of Sciences, Beijing 100049, China}
\affiliation{Key Laboratory of Radio Astronomy and Technology, Beijing 100101, China}
\email[show]{tww@bao.ac.cn}

\author[gname=Haiyan, sname=Zhang]{Haiyan Zhang}
\affiliation{National Astronomical Observatories, Chinese Academy of Sciences, Beijing 100012, China}
\affiliation{School of Astronomy and Space Science, University of Chinese Academy of Sciences, Beijing 100049, China}
\affiliation{Key Laboratory of Radio Astronomy and Technology, Beijing 100101, China}
\affiliation{Guizhou Radio Astronomical Observatory, Guizhou University, Guiyang 550000, China}
\email[show]{hyzhang@nao.cas.cn}

\author[gname=Dan, sname=Wu']{Dan Wu}
\affiliation{National Astronomical Observatories, Chinese Academy of Sciences, Beijing 100012, China}
\email{wudan@bao.ac.cn}


\begin{abstract}

We present a radio analysis of the recently identified supernova remnant G321.3–3.9 using archival multi-wavelength data spanning 88–2304 MHz. The source exhibits an elliptical shell-like morphology (1.3 $^{\circ}$ $\times$ 1.7 $^{\circ}$) and a relatively flat non-thermal spectral index of $\alpha$ = -0.40 $\pm$ 0.03.  The distance is estimated using both the $\Sigma$–$D$ relation (1.6–2.9 kpc) and tentative associations with HI structures, the latter suggesting a near-side solution of 2.5–3.3 kpc, though the physical connection remains uncertain.

\end{abstract}

\section{Introduction}
\label{sect:intro}

Supernova remnants (SNRs) are the expanding remains of stellar explosions, including core-collapse from massive stars and thermonuclear events from white dwarfs. Supernova explosions play a vital role in the interstellar medium (ISM) by injecting energy, enriching it with heavy elements, triggering star formation through shock waves, and serving as major accelerators of high-energy cosmic rays \citep{2020pesr.book.....V}. Understanding SNRs is essential for insights into stellar evolution, galactic dynamics, and high-energy astrophysics. Although the predicted Galactic supernova rate (about 2-3 per century) suggests that over 1000 SNRs should exist in the Galactic system, only about 300 are currently listed in the Green SNR Catalog \citep{2025JApA...46...14G}, primarily due to the limited sensitivity of available radio continuum data and confusion with other extended radio sources, especially HII regions.

High-sensitivity radio surveys are still the main way to discover those missing SNRs, and many SNR candidates have also been found in previous radio survey data, awaiting further analysis and confirmation. In this article, we focus on G321.3–3.9, which was previously classified as a SNR candidate and has recently been identified as a SNR by \citet{2024A&A...690A.278M}. G321.3–3.9 was first detected as a partial elliptical structure by \citet{1997MNRAS.287..722D} at 2.4 GHz using the Parkes 64 m telescope and later detected in the 843 MHz MGPS-2 survey \citep{2024A&A...690A.278M}. Recently, \citet{2024ApJS..272...36F} revealed significant optical line emissions in G321.3–3.9, with elliptical filamentary structures visible in both H$\alpha$ and [O III], but more clearly defined in the [O III] image, suggesting that these regions may have entered the radiative phase. \citet{2024A&A...690A.278M} analyzed observations of the remnant in radio and X-ray data, calculating a radio spectral index of $\alpha$ = -0.8$\pm$0.2. However, as the spectral index was derived from only two frequency points due to restricted baseline coverage, it may not fully capture the complete complexity of the radio spectrum.

In this work, we use archival, multi-wavelength radio data to analyze the properties of the newly identified SNR G321.3–3.9. The paper is organized as follows. Sect. \ref{sect:data} describes the data sets used in this study. Sect. \ref{sect:properties} summarizes the main properties of the remnant, including the radio spectral index, polarization features, and distance estimates. Sect. \ref{sect:analysis} discusses the results and their implications, and Sect. \ref{sect:conclusion} presents our conclusions.
\section{DATA}
\label{sect:data}

    \begin{figure}
    \centering
    \includegraphics[width=1.0\textwidth, angle=0]{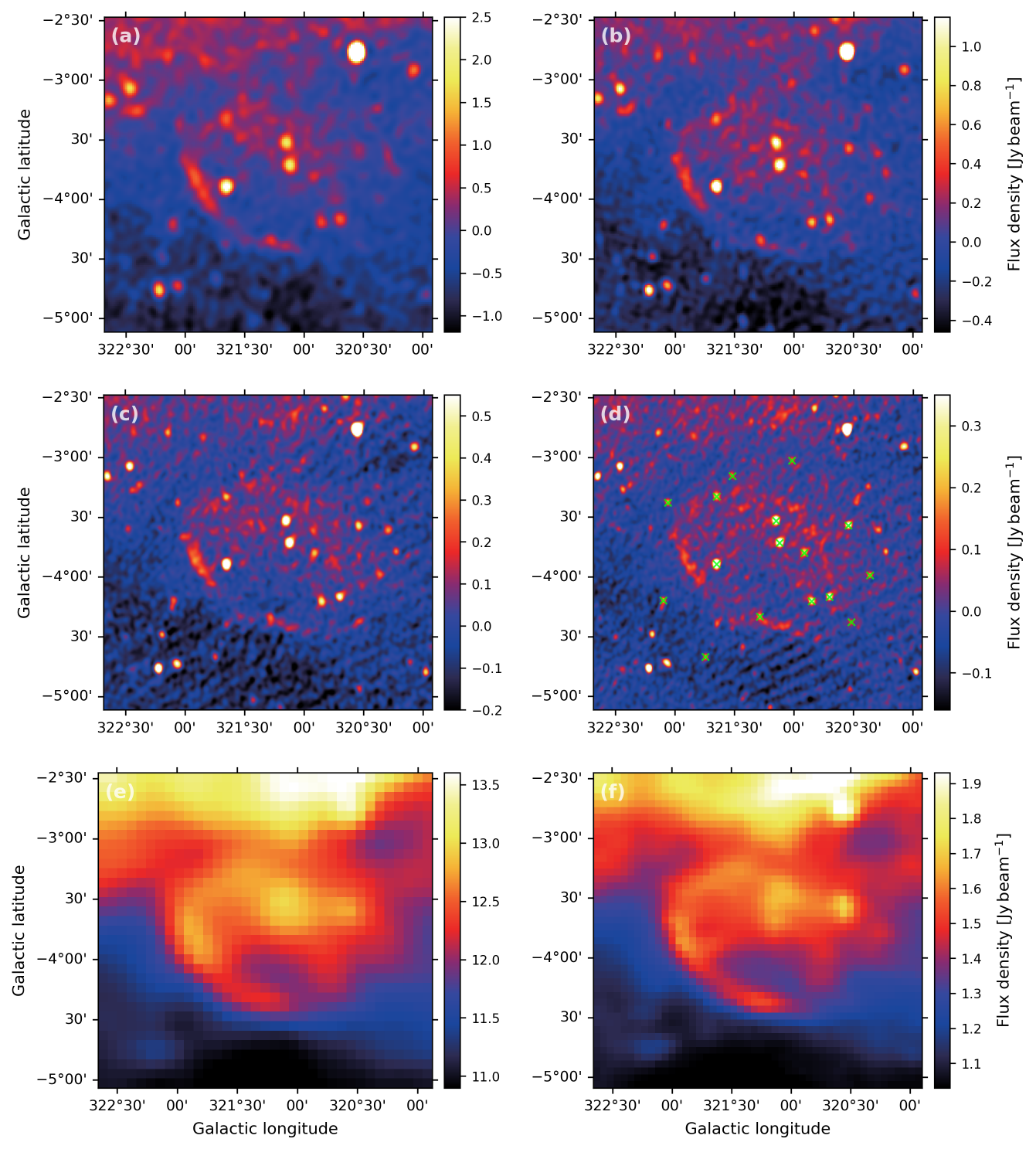}
    \caption{Radio continuum observations of SNR G321.3-3.9 across six frequency bands, (a) 88 MHz, (b) 118 MHz, (c) 154 MHz, (d) 200 MHz, (e) 1.4 GHz and (f) 2.3 GHz. The colour scales on the right side have a unit of Jy beam$^{-1}$. Specifically, subtracted point sources are indicated by green crosses in panel (d).}
    \label{Fig1}
    \end{figure}

The low-frequency radio continuum data are drawn from the Galactic and Extra-Galactic All-Sky MWA Survey (GLEAM), which was conducted with the Murchison Widefield Array (MWA) at frequencies between 72 and 231 MHz \citep{2015PASA...32...25W, 2019PASA...36...47H}. The survey provides full coverage of the sky south of declination +30 $^{\circ}$ with spatial resolutions between 2 - 5 arcmin, depending on the frequency. Four wideband images spanning the frequency ranges 72–103 MHz, 103–134 MHz, 139–170 MHz, and 170–231 MHz were analysed in this study, hereafter referred to by their nominal central frequencies: 88 MHz, 118 MHz, 154 MHz, and 200 MHz. The 1.4 GHz data from CHIPASS, which is a new 1.4 GHz continuum map reprocessed from archival observations of the HI Parkes All-Sky Survey (HIPASS) and the HI Zone of Avoidance (HIZOA) survey, covers the sky south of $\delta$ = +25 $^{\circ}$ \citep{2014PASA...31....7C}. The final map achieves an angular resolution of about 14.4 arcmin and a sensitivity of roughly 40 mK. The 2.3 GHz, Stokes I, Q, and U data come from the S-band Polarization All-Sky Survey (S-PASS) \citep{2019MNRAS.489.2330C}, conducted with the 64-meter Parkes radio telescope. The survey covers the southern sky at declinations below -1 $^{\circ}$ with an angular resolution of 8.9 arcmin. The mean sensitivity for Stokes Q and U is 0.81 mK, while the Stokes I data are limited by a confusion level of 9 mK. The CHIPASS and S-PASS Stokes I data were utilized in \citet{2024A&A...690A.278M} to derive the spectral index of G321.3–3.9.

We also use CO and neutral hydrogen (HI) emission data to search for atomic or molecular clouds potentially associated with the remnant. The HI emission data used in this study are from the HI4PI survey \citep{2016A&A...594A.116H}, which combines observations from the Effelsberg-Bonn HI Survey (EBHIS) and the Galactic All-Sky Survey (GASS) into a homogeneous, full-sky HI dataset. We specifically utilize the southern sky data originating from the GASS component, which was observed with the Parkes 64-m telescope. The GASS dataset was initially released in 2009 \citep{2009ApJS..181..398M}, with improved versions published in 2010 \citep{2010A&A...521A..17K} and 2015 \citep{2015A&A...578A..78K}, the latter is the version incorporated into the HI4PI data release. The final HI4PI dataset achieves an rms brightness temperature noise of approximately 43 mK per channel, and provides an angular resolution of 16.2 arcmin and a velocity resolution of 1.49 km$\,\mathrm{s}^{-1}$. The velocity coverage spans from –470 to 470 km$\,\mathrm{s}^{-1}$ in the Local Standard of Rest (LSR) frame, enabling the study of both Galactic and high-velocity HI structures. The CO data used in this study are from the composite survey by \citet{2001ApJ...547..792D}, which combines 37 individual surveys of the \({}^{12}\text{CO} (J = 1 - 0)\) transition at 115 GHz. The survey covers the Galactic plane and nearby regions with varying resolution: regions observed in full-resolution surveys had angular resolutions of 8.4–8.8 arcmin, while some high-latitude or peripheral areas were only sparsely sampled at a coarser resolution of $\sim$0.5 $^{\circ}$, with data interpolated onto a uniform grid. Spectral resolution and noise levels vary across the survey, with channel widths ranging from 0.26 to 1.3 km$\,\mathrm{s}^{-1}$ and typical rms values between 0.12 and 0.43 K.
\section{Properties}
\label{sect:properties}

\begin{table*}
\centering
\caption{Flux densities (in Jy) and derived spectral index ($\alpha$) are provided for all identified point sources to be subtracted, as well as for SNR G321.3-3.9. The symbol * denotes points estimated by extrapolating fitting results from other frequency bands.}
\label{tab:point_sources}
\renewcommand{\arraystretch}{1.1}
\begin{tabular}{lccccccc}
\hline
\textbf{Source} & 
\multicolumn{1}{c}{\textbf{88 MHz}} &
\multicolumn{1}{c}{\textbf{118 MHz}} &
\multicolumn{1}{c}{\textbf{154 MHz}} &
\multicolumn{1}{c}{\textbf{200 MHz}} &
\multicolumn{1}{c}{\textbf{1395 MHz}} &
\multicolumn{1}{c}{\textbf{2303 MHz}} &
\multicolumn{1}{c}{\textbf{Spectral Index}} \\
\hline

G320.5-3.6  & 0.78$\pm$0.04 & 0.64$\pm$0.03 & 0.52$\pm$0.03 & 0.51$\pm$0.03 & 0.169$\pm$0.025* & 0.129$\pm$0.005* & -0.54$\pm$0.08\\
G321.6-3.9  & 2.98$\pm$0.15 & 2.21$\pm$0.11 & 1.63$\pm$0.08 & 1.23$\pm$0.06 & 0.150$\pm$0.025* & 0.087$\pm$0.005* & -1.08$\pm$0.08\\
G321.1-3.7  & 2.14$\pm$0.11 & 1.56$\pm$0.08 & 1.19$\pm$0.06 & 0.97$\pm$0.05 & 0.146$\pm$0.025* & 0.090$\pm$0.005* & -0.96$\pm$0.08\\
G321.1-3.5  & 1.88$\pm$0.10 & 1.31$\pm$0.07 & 1.08$\pm$0.05 & 0.85$\pm$0.04 & 0.132$\pm$0.025* & 0.082$\pm$0.005* & -0.95$\pm$0.08\\
G320.7-4.2  & 1.11$\pm$0.06 & 0.80$\pm$0.04 & 0.70$\pm$0.04 & 0.56$\pm$0.03 & 0.118$\pm$0.025* & 0.079$\pm$0.005* & -0.80$\pm$0.08\\
G320.9-4.2  & 1.27$\pm$0.06 & 0.94$\pm$0.05 & 0.54$\pm$0.03 & 0.47$\pm$0.02 & 0.036$\pm$0.025* & 0.019$\pm$0.005* & -1.29$\pm$0.08\\
G322.1-3.4  & 0.62$\pm$0.03 & 0.31$\pm$0.02 & 0.25$\pm$0.01 & 0.27$\pm$0.01 & 0.032$\pm$0.025* & 0.020$\pm$0.005* & -1.00$\pm$0.08\\
G321.6-3.3  & 1.01$\pm$0.05 & 0.65$\pm$0.03 & 0.32$\pm$0.02 & 0.33$\pm$0.02 & 0.016$\pm$0.025* & 0.007$\pm$0.005* & -1.49$\pm$0.08\\
G321.3-4.3  & 1.32$\pm$0.07 & 0.83$\pm$0.04 & 0.47$\pm$0.02 & 0.24$\pm$0.01 & 0.004$\pm$0.025* & 0.002$\pm$0.005* & -2.09$\pm$0.08\\
G321.0-3.0  & 0.36$\pm$0.02 & 0.33$\pm$0.02 & 0.21$\pm$0.01 & 0.15$\pm$0.01 & 0.017$\pm$0.025* & 0.010$\pm$0.005* & -1.14$\pm$0.09\\
G321.7-4.7  & 0.67$\pm$0.03 & 0.47$\pm$0.02 & 0.35$\pm$0.02 & 0.32$\pm$0.02 & 0.047$\pm$0.025* & 0.029$\pm$0.005* & -0.94$\pm$0.08\\
G321.5-3.2  & 0.63$\pm$0.03 & 0.34$\pm$0.02 & 0.21$\pm$0.01 & 0.19$\pm$0.01 & 0.008$\pm$0.025* & 0.004$\pm$0.005* & -1.54$\pm$0.09\\
G322.1-4.2  & 1.22$\pm$0.06 & 0.68$\pm$0.04 & 0.41$\pm$0.02 & 0.33$\pm$0.02 & 0.013$\pm$0.025* & 0.006$\pm$0.005* & -1.62$\pm$0.08\\
G320.9-3.8  & 0.65$\pm$0.03 & 0.35$\pm$0.02 & 0.48$\pm$0.02 & 0.37$\pm$0.02 & 0.142$\pm$0.025* & 0.111$\pm$0.005* & -0.49$\pm$0.08\\
G320.4-4.0  & 0.61$\pm$0.03 & 0.49$\pm$0.02 & 0.42$\pm$0.02 & 0.31$\pm$0.02 & 0.069$\pm$0.025* & 0.046$\pm$0.005* & -0.79$\pm$0.08\\
G320.5-4.4  & 0.46$\pm$0.02 & 0.24$\pm$0.01 & 0.29$\pm$0.02 & 0.13$\pm$0.01 & 0.013$\pm$0.025* & 0.007$\pm$0.005* & -1.29$\pm$0.09\\

\hline
G321.3-3.9  & 56.68$\pm$5.67 & 48.05$\pm$4.81 & 38.41$\pm$3.84 & 32.72$\pm$3.27 & 17.67$\pm$1.77 & 13.66$\pm$1.37 & -0.40$\pm$0.03\\
\hline
\end{tabular}
\end{table*}

\begin{figure*}
\gridline{
    \includegraphics[width=0.46\textwidth]{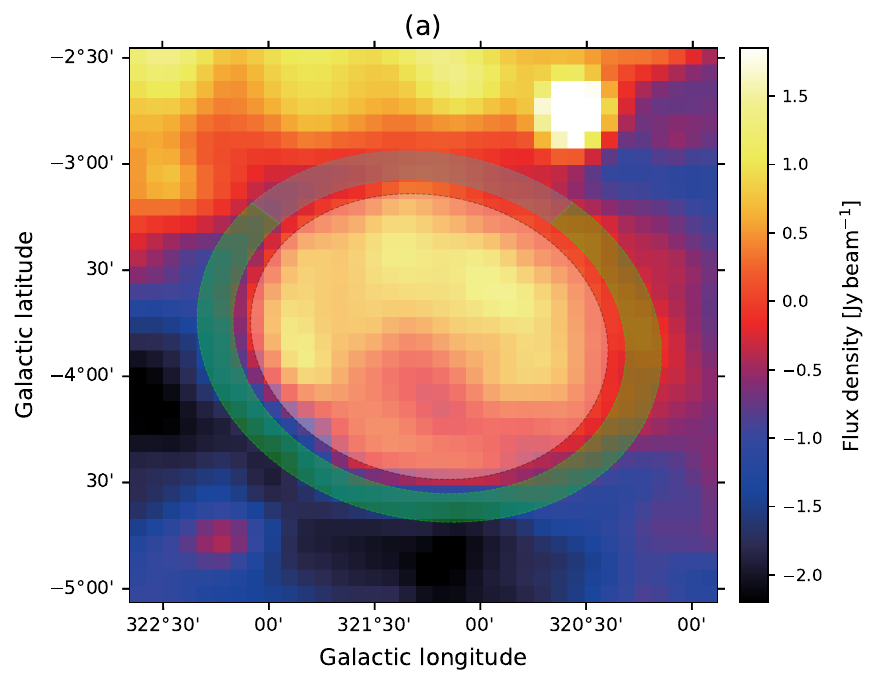}\hfill
    \includegraphics[width=0.54\textwidth]{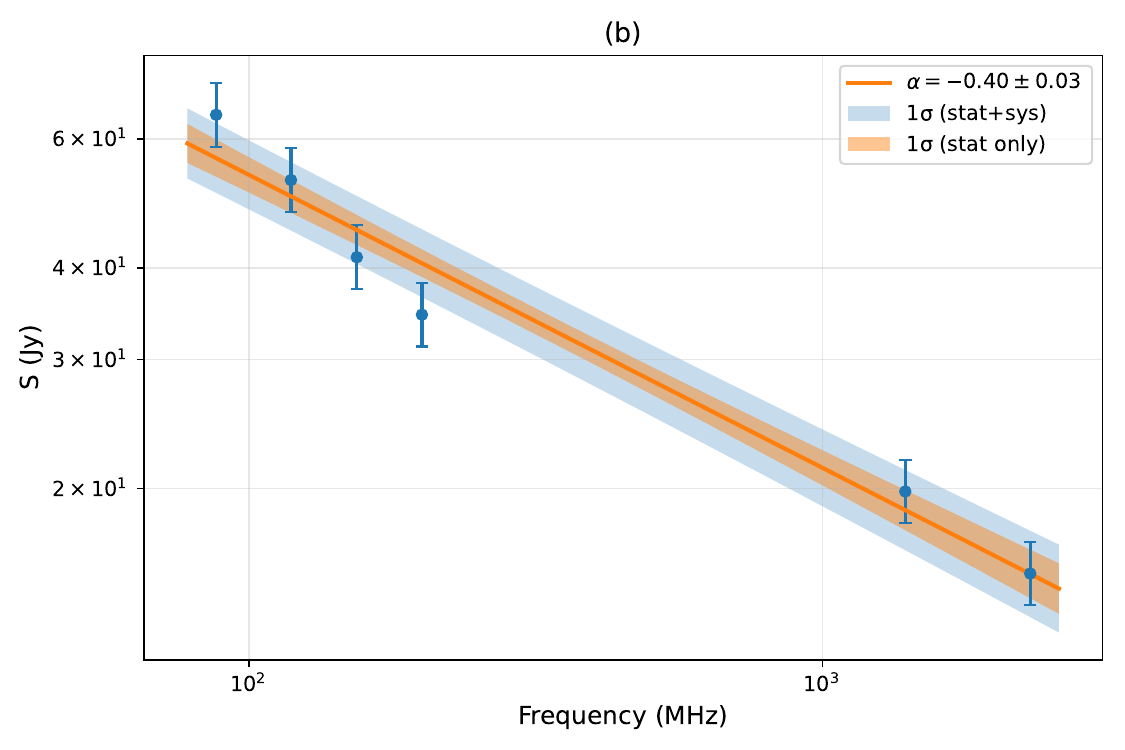}
}
\caption{(a) 200 MHz intensity map after point source subtraction, convolved to a 14.4' beam and regridded to the 1.4 GHz pixel scale. The source region is shown as the central white ellipse, the background as the surrounding green annulus, while the northern gray area is excluded to avoid contamination. (b) Integrated spectrum of SNR G321.3-3.9. Blue points represent the mean flux densities with statistical uncertainties. The orange line shows the best-fit power-law model, with shaded bands for statistical (orange) and total (blue) uncertainties.}
\label{Fig2}
\end{figure*}

\subsection{Radio spectral index}

 As shown in Figure~\ref{Fig1}, SNR G321.3–3.9 exhibits similar morphological features across multiple radio wavelengths. It presents a clear elliptical shell structure with an angular size of 1.3 $^{\circ}$ × 1.7 $^{\circ}$, which suggests interactions with the non-uniform surrounding ISM. At lower frequencies, the surrounding environment appears relatively clean without noticeable extended structures. However, in the 1.4 GHz and 2.3 GHz data, which have larger beam sizes, a bright extended structure is observed to the north of the remnant.

 The first step in the processing procedure is to subtract the contribution of unrelated point sources to ensure accurate flux measurements. Using a method similar to \citet{2005A&A...436..187T}, we selected a small area around each point source and applied two-dimensional Gaussian fitting to eliminate their influence. Combining multi-wavelength radio images with prior source catalogs (e.g., \citealt{2021PASA...38...58H}), we determined 19 point sources to be subtracted, all of which are listed in Table \ref{tab:point_sources} and marked with green crosses in Figure~\ref{Fig1}d. Due to resolution limitations, point sources in the 1.4 GHz and 2.3 GHz data could not be individually identified. Instead, their flux contributions were estimated by extrapolating fitting results from other frequency bands and were subsequently subtracted after convolution to the corresponding resolution.

Once these corrections were completed, we proceeded to convolve all data to a common beam and re-grid them onto the same spatial grid, using the lowest-resolution 1.4 GHz data as reference to minimize additional errors. For the processed images, we adopted consistent source and background regions, as illustrated in Fig.~\ref{Fig2}a: the source region is the white elliptical area, while the background is an elliptical annulus shown in green, excluding the gray portion to avoid contamination from northern extended structures. Based on the flux densities measured within these regions, we fitted a power-law spectrum to derive the spectral index, and further assessed its associated uncertainties.

Both statistical and systematic uncertainties need to be taken into account. The statistical uncertainty of the spectral index originates from the flux measurement noise, while the systematic component reflects methodological effects. Given the complex surroundings of the remnant, the choice of background region is assumed as the dominant source of systematics. To quantify this effect, we extracted fluxes using nine alternative background definitions, and constructed the sample covariance of the resulting flux vectors (all background regions are shown in Appendix Fig.~\ref{apfig1}). Using generalized least squares with this covariance, We derived a best-fit spectral index of $\alpha = -0.40 \pm 0.03_{\rm stat} \pm 0.01_{\rm sys}$, corresponding to a quadrature sum of $\pm 0.03$ for the total uncertainty, as shown in Fig.~\ref{Fig2}b. The inner band reflects statistical uncertainties only, while the outer band shows the total uncertainty, including the contribution from systematic component.

\begin{figure}
\centering
\includegraphics[width=0.75\textwidth, angle=0]{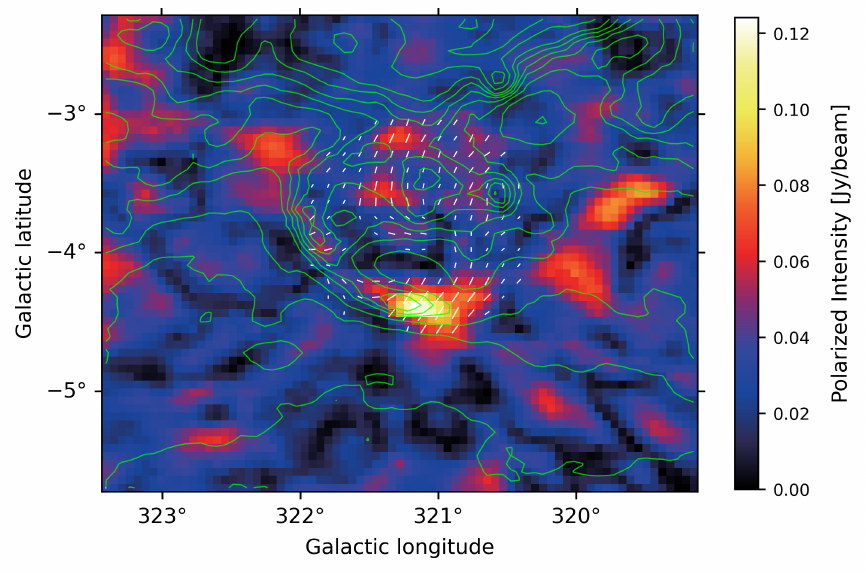}
\caption{Polarized intensity image of G321.3–3.9 at 2.3 GHz from S-PASS. White lines indicate the orientation of B-vectors, with lengths proportional to the local polarized intensity. Green contours show the continuum intensity at the same frequency, also derived from S-PASS.}
\label{Fig3}
\end{figure}

\subsection{Polarization}
Radio polarization measurements of SNRs are useful for probing the magnetic field structure, as synchrotron radiation is the dominant emission mechanism, but the inferred magnetic field is limited by projection effects and observational uncertainties. We obtained the PI (polarized intensity, $PI= \sqrt{Q^2 + U^2- \sigma^2}$ ) image of G321.3–3.9 at 2.3 GHz using the Stokes Q, U, and sensitivity data from the 2.3 GHz S-PASS survey, as shown in Fig. \ref{Fig3}. Additionally, the B-vector around the remnant is indicated with overlaid white arrows, representing the projected direction of the local magnetic field, with arrow lengths proportional to the polarized intensity.

The observed B-vector around the remnant are generally tangential and exhibits a circular pattern near the center, possibly due to compression and amplification by the explosion shock wave. Additionally, the strongest polarized emission is observed near the southern boundary of G321.3–3.9, where the polarization fraction reaches approximately 10\%. Notably, the spatial coincidence with the remnant may not originate from the remnant itself, especially given the difference between the polarized and total intensity distributions. Moreover, the analysis of the magnetic field is strongly affected by the characteristics of the remnant, such as its age, size, and surrounding environment, making it difficult to draw firm conclusions based solely on polarized intensity


\begin{figure}
\centering
\includegraphics[width=1.0\textwidth, angle=0]{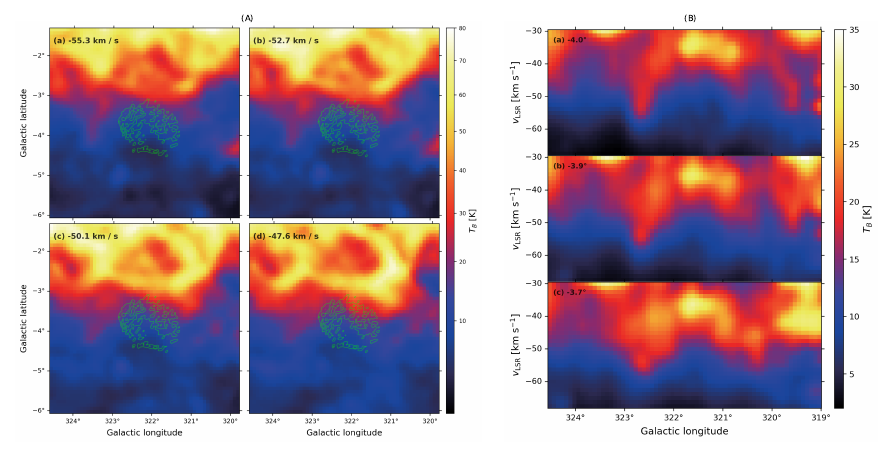}
\caption{(A) HI brightness temperature maps around SNR G321.3–3.9, averaged over velocity channels of 2.5 km\,s$^{-1}$ from $-55$ to $-47$ km\,s$^{-1}$. Central LSR velocities are labeled in each panel. Green contours show the radio continuum intensity at 200 MHz, outlining the morphology of the remnant. (B) HI position–velocity diagrams toward the remnant, extracted as glon–velocity slices from the HI cube with a latitude width of 0.2$^{\circ}$. Central latitudes are indicated in each panel.}
\label{Fig4}
\end{figure}

\subsection{Distance Estimate}

For SNR G321.3$-$3.9, we applied two approaches to estimate its distance: one based on the empirical $\Sigma$--$D$ relation, and the other on potential interstellar gas structures physically associated with the remnant. In the $\Sigma$--$D$ method, the radio surface brightness $\Sigma$ and the source diameter $D$ are related by $\Sigma = A D^{-\beta}$. The surface brightness $\Sigma$ at 1\,GHz was derived from our fitted flux density and, using the simulation--filter calibration of \citet{2019SerAJ.199...23S}, this corresponds to a physical diameter of about 56.9\,pc. Together with the equivalent angular diameter of the remnant, this yields a representative distance of $\sim$2.2\,kpc. Taking into account both the propagated measurement errors and the intrinsic scatter of the $\Sigma$--$D$ relation, this results in a plausible range of 1.6--2.9\,kpc.

Another approach to distance estimation relies on identifying molecular or neutral gas structures associated with the remnant. Their radial velocities can then be translated into kinematic distances using the Galactic rotation model. Since SNRs are often observed in contact with molecular or neutral hydrogen clouds \citep{2010ApJ...712.1147J,2014IAUS..296..170C,2023ApJS..268...61Z}, we examined the HI and $^{12}$CO (J=1–0) data cubes around SNR G321.3$-$3.9 to search for such potential associations. In the CO data, no emission features with a signal-to-noise ratio above 5 were detected toward the remnant. However, the region was only sparsely sampled with a beam size of $\sim$0.5 $^{\circ}$, which may have smeared out compact molecular emission below the detection limit, and thus the possibility of associated molecular clouds cannot be ruled out.

In the HI data, we examined position–position (PP) and position–velocity (PV) maps around the location of the remnant and identified tentative associated features. The PP maps at velocity slices between -47 and -55 km s$^{-1}$ are shown in Fig.~\ref{Fig4}a, where the HI emission is concentrated mainly in the northern part of the field. The central region of the remnant coincides with a local brightness depression, while its northern boundary is adjacent to a region of enhanced HI emission. Three PV maps extracted along Galactic latitudes from $b=-3.7^{\circ}$ to $-4.0^{\circ}$ are presented in Fig.~\ref{Fig4}b. These reveal a partial arc-like structure, roughly centered at Galactic longitude $l=321.5^{\circ}$ and velocity $v_{\rm LSR}\approx -50$\,km\,s$^{-1}$. Following the kinematic distance method of \citet{2018ApJ...856...52W}, which employs the Galactic rotation curves calibrated by VLBI parallaxes \citep{2014ApJ...783..130R,2019ApJ...885..131R}, we adopt $v_{\rm LSR}\approx-50$ km s$^{-1}$ as the approximate systemic velocity for the HI feature, with an estimated measurement uncertainty of $\pm 4$ km s$^{-1}$ inferred from its positional ambiguity in the PP/PV maps, to estimate the distance. The resulting near-side and far-side distances are 2.9 and 9.9 kpc, with 68\% credible intervals (CIs) of 2.5–3.3 and 9.5–10.3 kpc, respectively, obtained through its Monte Carlo implementation, which accounts for uncertainties in both the corrected $v_{\rm LSR}$ (including solar motion and the assumed velocity measurement uncertainty) and the Galactic rotation-curve parameters. It should be noted that the distance estimation refers only to the HI feature itself, and its possible association with the remnant remains tentative pending further evidence.

\section{Discussion}
\label{sect:analysis}

In radio surveys, the radio spectral index $\alpha$ is commonly used to distinguish supernova remnants from other extended sources. The average spectral index of shell-type SNRs is approximately -0.5. By contrast, HII regions typically have a spectral index of around -0.1, while pulsar wind nebulae exhibit slightly flatter spectra, usually ranging from -0.2 to -0.3 \citep{2020pesr.book.....V}. In addition, the radio spectral index distribution of SNRs is quite broad, influenced by factors such as age, ambient medium, and observational uncertainties. The relatively flat spectral index of G321.3–3.9 is consistent with a more evolved SNR, as younger remnants tend to have steeper spectra. The physical origin of such spectral flattening is not yet fully understood, but several mechanisms have been proposed, including reduced shock acceleration efficiency, turbulent reacceleration in the post-shock region, and high post-shock compression ratios associated with radiative shocks \citep{2020pesr.book.....V}. Regarding polarization, the B-vectors are broadly tangential, but the polarized emission does not closely trace the total-intensity shell. Given possible depolarization effects at 2.3 GHz, no firm conclusions can be drawn from the present data.

Estimating the distance of the remnant is essential for assessing its physical size and evolutionary state, and we therefore compare several independent approaches whose ranges show limited consistency and warrant further discussion. From the $\Sigma$--$D$ relation we infer a distance of 1.6–2.9 kpc. This method provides a practical estimate based on the remnant’s surface brightness, although its reliability depends on the adopted calibration as well as on the remnant’s intrinsic properties and environment. Given the uncertain association between the HI structure and the remnant, we treat the kinematic distance inferred from the rotation curve as a reference estimate only. The near-side solution, centered at $\sim$2.8 kpc, lies close to the upper bound of the $\Sigma$--$D$ range, where the kinematic and $\Sigma$--$D$ estimates converge, whereas the far-side solution is excluded, as it would imply a physical size implausibly large for a typical SNR. For comparison, \citet{2024A&A...690A.278M} established broad lower and upper bounds on the distance to G321.3–3.9 ($\sim$0.7–3.4 kpc) from radio surface brightness, and further derived a distance of 1.0–1.7 kpc by combining X-ray absorption with extinction data. Our results fall within their broader bounds, though only the $\Sigma$--$D$ based value partially overlaps with their estimate. This discrepancy may partly reflect systematic differences between methods, suggesting that the distance to the remnant remains uncertain and will require further observational constraints.

\section{Conclusion}
\label{sect:conclusion}

We presented a radio analysis of the recently identified SNR G321.3–3.9 using archival multi-wavelength radio data. The remnant exhibits a clear elliptical shell-like morphology. The integrated spectrum yields a spectral index of $\alpha = -0.40 \pm 0.03$, where the quoted uncertainty reflects both statistical and systematic contributions. Polarization at 2.3\,GHz points to broadly tangential B-vectors but remains inconclusive owing to projection and depolarization. For the distance, we applied both the $\Sigma$–$D$ relation and potential associations with interstellar gas structures, and compared the results with previous studies. The different methods yield partially inconsistent ranges, highlighting significant systematic uncertainties in the distance measurement. Future observations of this remnant will be valuable for clarifying these uncertainties. Higher-resolution CO and HI data, together with X-ray measurements, will be important for refining the distance and characterizing the surrounding interstellar medium, while broadband radio polarization will help to constrain the magnetic-field structure. In addition, hydrodynamical simulations that include magnetic fields and a non-uniform ambient medium may provide a more realistic picture of the remnant’s evolution and morphology.

\begin{acknowledgments}

This work is partially supported by the National SKA Program of China (2025SKA0140100), the National Key R\&D Program of China (No.2023YFA1608000), the Youth Innovation Promotion Association of CAS (2023000015), the China Manned Space Program with grant No.CMS-CSST-2025-A14, the National Natural Science Foundation of China (No.12041301), and the Guizhou Provincial Science and Technology Projects (No.QKHFQ[2023]003, No.QKHFQ[2024]001, No.QKHPTRC-ZDSYS[2023]003). The computing task was carried out on the Science Platform at China National Astronomical Data Center (NADC). NADC is a National Science and Technology Innovation Base hosted at National Astronomical Observatories, Chinese Academy of Sciences. 
\end{acknowledgments}





%
\facilities{MWA, Parkes, CfA:1.2m,}

\software{Astropy \citep{2013A&A...558A..33A,2018AJ....156..123A,2022ApJ...935..167A},
          Matplotlib \citep{team_matplotlib_2024},
          NumPy \citep{harris_array_2020},
          pandas \citep{team_pandas-devpandas_2024},
          regions \citep{bradley_astropyregions_2024},
          reproject \citep{robitaille_astropyreproject_2024},
          SciPy \citep{virtanen_scipy_2020},
          spectral-cube \citep{ginsburg_radio-astro-toolsspectral-cube_2019},
          kd \citep{trey_wenger_2021_5660905},}


\appendix

\section{Alternative background selections}
\label{appendix1}

\begin{figure}
\centering
\includegraphics[width=0.75\textwidth, angle=0]{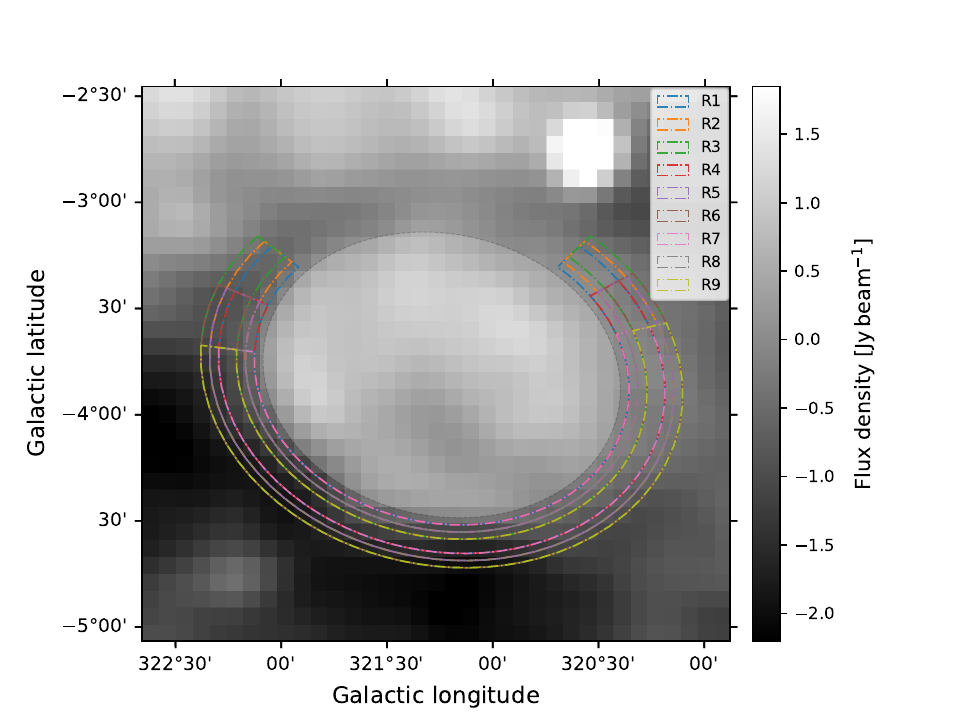}
\caption{Source region (white) and nine alternative background regions (R1–R9) adopted to assess systematic uncertainties in the spectral index. The region labeled R3 corresponds to the example background shown in Figure \ref{Fig2}, while the spectral analysis itself was performed using the full set of regions.}
\label{apfig1}
\end{figure}

To assess the systematic uncertainty introduced by the choice of background, we tested nine alternative elliptical annuli, defined by varying the inner and outer radii relative to the source ellipse, as well as the opening angle. The full set of regions is shown in Figure \ref{apfig1}, where the source region is marked at the center. Fluxes from these regions were used to construct the covariance matrix for the spectral index fitting.

\clearpage
\bibliography{references}{}
\bibliographystyle{aasjournalv7}



\end{document}